# Scattering and bound states for a class of non-central potentials

A. D. Alhaidari

*Physics Department, King Fahd University of Petroleum & Minerals, Dhahran 31261, Saudi Arabia*
e-mail: haidari@mailaps.org

We obtain $L^2$-series solutions of the nonrelativistic three-dimensional wave equation for a large class of non-central potentials that includes, as special cases, the Aharonov-Bohm, Hartmann, and magnetic monopole potentials. It also includes contributions from the potential term, $\cos\theta/r^2$ (in spherical coordinates). The solutions obtained are for all energies, the discrete (for bound states) as well as the continuous (for scattering states). The $L^2$ bases of the solution space are chosen such that the matrix representation of the wave operator is tridiagonal. The expansion coefficients of the radial and angular components of the wavefunction are written in terms of orthogonal polynomials satisfying three-term recursion relations resulting from the matrix wave equation.

## 1. Introduction

By relaxing the constraint of a diagonal representation of the Hamiltonian and allowing for the next higher level of generalization, which is that of tridiagonal matrices, we found a larger solution space for the wave equation with an extended class of exactly solvable potentials [1]. The $L^2$ series solutions obtained as such include the discrete (for bound states) as well as the continuous (for scattering states) spectrum of the Hamiltonian. Due to the tridiagonal structure of the matrix wave equation the problem, in this approach, translates into finding solutions of the resulting three-term recursion relation for the expansion coefficients of the wavefunction. These are written in terms of orthogonal polynomials, some of which are well-known but some are new while others are modified versions of known polynomials. In a recent article [1], we obtained solutions of problems in one and three dimensions using this approach. The solutions of some of the classic problems such as the Coulomb and Morse were reproduced adding, however, new tridiagonal representations to the solution space. We also found generalizations of others, such as the Hulthén problem, where we obtained an extended class of solutions and introduced their associated orthogonal polynomials. This kind of development embodies powerful tools in the analysis of solutions of the wave equation by exploiting the intimate connection and interplay between tridiagonal matrices and the theory of orthogonal polynomials. In such analysis, one is at liberty to employ a wide range of well established methods and numerical techniques associated with these settings such as quadrature approximation and continued fractions. These formulations were also extended to the study of the relativistic problem. The Dirac-Coulomb and Dirac-Morse are two relativistic problems, beside others, that have already been worked out using this approach [2].

In this article, we investigate the nonrelativistic problem in three dimensions with non-central potentials using the same approach and obtain scattering and bound states solutions of the wave equation. We consider the time-independent potential of the form $V(\vec{r}) = V(r,\theta)$ such that it is separable in spherical coordinates. Specifically, we study the following class of potentials

$$V(r,\theta) = V(r) + \frac{1}{2r^2}\left(\frac{\hat{C} + C\cos\theta}{\sin^2\theta} - C_0\cos\theta\right), \tag{1.1}$$

where $\hat{C}$, $C$, and $C_0$ are real potential parameters. The Aharonov-Bohm [3] and Hartmann [4] potentials are special cases for which $C = C_0 = 0$ (for pure Aharonov-Bohm effect, $\hat{C}$ is discrete via its linear dependence on an integer $m \in \mathbb{Z}$ which comes from the phase quanta of the angular component of the wavefunction, $e^{im\phi}$). The case where $C = \pm\hat{C}$ and $C_0 = 0$ corresponds to the magnetic monopole potential with singularity along the $\pm z$ axis [5]. Our main contribution to the solution of this kind of problems is two-fold. The first is the introduction of the three-dimensional potential term $\frac{\cos\theta}{r^2}$ which, to the best of our knowledge, was not treated exactly before. The second is the *simultaneous analytic* solution of scattering and bound states in the same formulation. We take the Coulomb interaction as the radial component of the potential. That is, we take $V(r) = \mathcal{Z}/r$, where $\mathcal{Z}$ is the electric charge coupling. Additionally, we consider briefly in Sec. 7 the radial oscillator potential $V(r) = \frac{1}{2}\omega^4 r^2$, where $\omega$ is the oscillator frequency.

For an introduction to the above-mentioned approach and its implementation on some examples in one and three dimensions (with spherical symmetry) one may consult the papers in Refs. [1,2]. Nonetheless, it might be useful to give, in few lines, a brief account as follows. Let $\{\varphi_n(\vec{r})\}_{n=0}^{\infty}$ be a complete set of $L^2$ basis in the configuration space with coordinates $\vec{r}$ that supports a tridiagonal matrix representation for the wave operator. That is, by expanding the wave function as $|\psi(\vec{r},E)\rangle = \sum_n f_n(E)|\varphi_n(\vec{r})\rangle$, the matrix representation of the wave operator in this basis could be written as follows

$$\langle\varphi_n|H - E|\varphi_m\rangle = (a_n - z)\delta_{n,m} + b_n\delta_{n,m-1} + b_{n-1}\delta_{n,m+1}, \tag{1.2}$$

where $z$ and the coefficients $\{a_n, b_n\}_{n=0}^{\infty}$ are real and, in general, functions of the energy, $E$, angular momentum, and potential parameters. Therefore, the matrix wave equation, which is obtained by expanding $|\psi\rangle$ in $(H - E)|\psi\rangle = 0$ as $\sum_m f_m|\varphi_m\rangle$ and projecting on the left by $\langle\varphi_n|$, results in the following three-term recursion relation

$$z f_n = a_n f_n + b_{n-1}f_{n-1} + b_n f_{n+1}. \tag{1.3}$$

Consequently, the problem translates into finding solutions of this recursion relation for the expansion coefficients of the wavefunction. In most cases this recurrence relation could be solved easily and directly by correspondence with those for well known orthogonal polynomials. Moreover, Eq. (1.2) shows that the discrete energy spectrum is easily obtained by imposing the diagonalization constraint which requires that

$$b_n = 0, \quad a_n - z = 0. \tag{1.4}$$

for all $n$.

We start in the following section by formulating the problem and writing down the basis elements for the angular and radial components of the wavefunction that support a tridiagonal matrix representation for the associated wave operator. The corresponding components of the non-central separable potential that are compatible with the tri-diagonal representations are also obtained. Explicit construction of the angular wave function for three possible configurations is given in Sec. 3. The radial component, which is compatible with the Coulomb interaction, is obtained in Sc. 4. The complete solution space splits into two disconnected subspaces: one for $C_0 \neq 0$ and another for $C_0 = 0$. These are constructed in Sec. 5 and Sec. 6, respectively. Additionally, in Sec. 5, we give the complete and explicit solution for the special case where $\hat{C} = C = 0$ and $C_0 \neq 0$, which is unique to the present work. In Sec. 7, we study briefly the case where the radial component of the non-central potential is that of the spherical oscillator. The new orthogonal polynomials associated with the tridiagonal representation of the angular component of the solution space will be investigated in Appendix B. The resolvent operator and weight (density) function for these polynomials are obtained in terms of the recursion coefficients $\{a_n, b_n\}$. Finally, in Appendix C we formulate the problem of a charged particle moving in a cylindrical electromagnetic vector potential (e.g., outside an infinitely long and thin current solenoid) and establish its connection to the present problem. As an example, we obtain the bound states solution for the combined Aharonov-Bohm effect and a magnetic monopole.

## 2. Non-central separable potentials in spherical coordinates

In the atomic units $\hbar = m = 1$, the time-independent Schrödinger wave equation for a structureless scalar particle of mass $m$ in a potential $V(\vec{r})$ is

$$\left[ -\frac{1}{2} \vec{\nabla}^2 + V(\vec{r}) - E \right] \psi = 0 , \qquad (2.1)$$

where $\vec{\nabla}$ is the three-dimensional Laplacian. The energy, $E$, is real and it is either discrete for bound states, or continuous for scattering states. In the spherical coordinates, $\vec{r} = \{r, \theta, \phi\}$, this wave equation could be written explicitly as follows

$$\left\{ \frac{1}{r^2} \frac{\partial}{\partial r} r^2 \frac{\partial}{\partial r} + \frac{1}{r^2} \left[ (1-x^2) \frac{\partial^2}{\partial x^2} - 2x \frac{\partial}{\partial x} + \frac{1}{1-x^2} \frac{\partial^2}{\partial \phi^2} \right] - 2V + 2E \right\} \psi = 0 , \qquad (2.2)$$

where $x = \cos\theta$. Consequently, this equation is separable for potentials of the form

$$V(\vec{r}) = V_r(r) + \frac{1}{r^2} \left[ V_\theta(x) + \frac{1}{1-x^2} V_\phi(\phi) \right] . \qquad (2.3)$$

This is so because if we write the wavefunction as $\psi(r, \theta, \phi) = r^{-1} R(r) \Theta(\theta) \Phi(\phi)$, then the wave equation (2.2) with the potential (2.3) gets separated in all three coordinates as follows

$$\left( \frac{d^2}{d\phi^2} - 2V_\phi + 2E_\phi \right) \Phi = 0 , \qquad (2.4a)$$

$$\left[ (1-x^2) \frac{d^2}{dx^2} - 2x \frac{d}{dx} - \frac{2E_\phi}{1-x^2} - 2V_\theta + 2E_\theta \right] \Theta = 0 , \qquad (2.4b)$$

$$\left( \frac{d^2}{dr^2} - \frac{2E_\theta}{r^2} - 2V_r + 2E \right) R = 0 , \qquad (2.4c)$$

where $E_\phi$ and $E_\theta$ are the separation constants, which are real and dimensionless. Square integrability of the basis is with respect to the following integration measures

$$\int |\psi|^2 d^3\vec{r} = \int_0^\infty |R|^2 dr \int_{-1}^{+1} |\Theta|^2 dx \int_0^{2\pi} |\Phi|^2 d\phi. \tag{2.5}$$

They are also required to satisfy the boundary conditions that $R(0) = R(\infty) = 0$, $\Phi(\phi) = \Phi(\phi + 2\pi)$, $\Theta(0)$ and $\Theta(\pi)$ are finite. If we specialize to the case where $V_\phi = 0$, then the normalized solution of Eq. (2.4a) that satisfies the boundary conditions is

$$\Phi_m(\phi) = \frac{1}{\sqrt{2\pi}} e^{im\phi}, \quad m = 0, \pm 1, \pm 2, \ldots, \tag{2.6}$$

giving $E_\phi = \frac{1}{2} m^2$.

The $L^2$ basis elements for the angular wavefunction component $\Theta(\theta)$ in the configuration space with coordinate $x \in [-1, +1]$ that satisfy the boundary conditions are

$$\chi_n(x; \mu, \nu) = A_n (1-x)^\alpha (1+x)^\beta P_n^{(\mu,\nu)}(x), \tag{2.7}$$

where $x = \cos\theta$, $P_n^{(\mu,\nu)}(x)$ is the Jacobi polynomial of order $n$ and $n = 0, 1, 2, \ldots$ The dimensionless real parameters $\alpha, \beta > 0$, $\mu, \nu > -1$ and $A_n$ is the normalization constant

$$A_n = \sqrt{\frac{2n + \mu + \nu + 1}{2^{\mu+\nu+1}} \frac{\Gamma(n+1)\Gamma(n+\mu+\nu+1)}{\Gamma(n+\mu+1)\Gamma(n+\nu+1)}}. \tag{2.8}$$

Using the differential equation (A.3) and differential formula (A.4) for the Jacobi polynomials, shown in Appendix A, we obtain

$$\left[(1-x^2)\frac{d^2}{dx^2} - 2x\frac{d}{dx}\right]\chi_n = \left[-n\left(x + \frac{\nu-\mu}{2n+\mu+\nu}\right)\left(\frac{\mu-2\alpha}{1-x} + \frac{2\beta-\nu}{1+x}\right) + \alpha^2 \frac{1+x}{1-x} + \beta^2 \frac{1-x}{1+x}\right.$$
$$\left. -(2\alpha\beta + \alpha + \beta) - n(n+\mu+\nu+1)\right]\chi_n + 2\frac{(n+\mu)(n+\nu)}{2n+\mu+\nu}\left(\frac{\mu-2\alpha}{1-x} + \frac{2\beta-\nu}{1+x}\right)\frac{A_n}{A_{n-1}} \chi_{n-1} \tag{2.9}$$

Therefore, the action of the differential operator of Eq. (2.4b) on the basis element (2.7) reads as follows

$$(H_\theta - E_\theta)\chi_n = \left[\frac{n}{2}\left(x + \frac{\nu-\mu}{2n+\mu+\nu}\right)\left(\frac{\mu-2\alpha}{1-x} + \frac{2\beta-\nu}{1+x}\right) - \frac{\alpha^2}{2}\frac{1+x}{1-x} - \frac{\beta^2}{2}\frac{1-x}{1+x} + \frac{E_\phi}{1-x^2} + V_\theta\right.$$
$$\left. -E_\theta + \alpha\beta + \frac{\alpha+\beta}{2} + \frac{n}{2}(n+\mu+\nu+1)\right]\chi_n - \frac{(n+\mu)(n+\nu)}{2n+\mu+\nu}\left(\frac{\mu-2\alpha}{1-x} + \frac{2\beta-\nu}{1+x}\right)\frac{A_n}{A_{n-1}} \chi_{n-1} \tag{2.10}$$

The recurrence relation (A.1) and orthogonality relation (A.5) of the Jacobi polynomials show that a tridiagonal matrix representation, $\langle \chi_n | H_\theta - E_\theta | \chi_{n'} \rangle$, is obtainable only for a limited number of special choices of potential components $V_\theta$ and for specific relations among the parameters as follows:

(1) $\alpha = \frac{\mu}{2}$, $\beta = \frac{\nu}{2}$, and $V_\theta = \frac{1}{2} \frac{\frac{1}{2}(\mu^2 + \nu^2) - m^2 + \frac{1}{2}(\mu^2 - \nu^2)x}{1 - x^2} - \frac{C_0}{2} x$ (2.11a)

(2) $\alpha = \frac{\mu}{2}$, $\beta = \frac{\nu+1}{2}$, and $V_\theta = \frac{1}{2} \frac{\frac{1}{2}(\mu^2 + C_1) - m^2 + \frac{1}{2}(\mu^2 - C_1)x}{1 - x^2}$ (2.11b)

(3) $\alpha = \frac{\mu+1}{2}$, $\beta = \frac{\nu}{2}$, and $V_\theta = \frac{1}{2} \frac{\frac{1}{2}(C_2 + \nu^2) - m^2 + \frac{1}{2}(C_2 - \nu^2)x}{1 - x^2}$ (2.11c)

where $m = 0, \pm 1, \pm 2,..$ and $\{C_i\}_{i=0}^{2}$ are dimensionless real parameters. The first possibility (2.11a) eliminates the $\chi_{n-1}$ term from Eq. (2.10), whereas the last two allow this term to contribute to the matrix elements above and below the diagonal. The details of these three possibilities will be given in the following section.

Now, the radial component of the wavefunction, $R(r)$, could be taken as an element in the space spanned by the $L^2$ functions

$$\xi_n(y; \lambda, \nu) = B_n y^\alpha e^{-y/2} L_n^\nu(y), \tag{2.12}$$

where $y = \lambda r$ and $L_n^\nu(y)$ is the Laguerre polynomial of order $n$. The real parameter $\lambda$ is positive and carries the dimension of inverse length (i.e., it is a length scale parameter). On the other hand, the dimensionless parameters $\alpha > 0$ and $\nu > -1$. It should be understood that the basis parameters $\alpha$ and $\nu$ are reused here for the purpose of economy in the use of symbols but are not the same as those that appear in the angular wave function basis (2.7). The normalization constant $B_n = \sqrt{\lambda \Gamma(n+1)/\Gamma(n+\nu+1)}$. Using the differential equation (A.8) and differential formula (A.9) for the Laguerre polynomials, we obtain

$$\begin{aligned}\frac{d^2 \xi_n}{dr^2} = \lambda^2 &\left[ -\frac{n}{y}\left(1 + \frac{\nu+1-2\alpha}{y}\right) + \frac{\alpha(\alpha-1)}{y^2} - \frac{\alpha}{y} + \frac{1}{4} \right] \xi_n \\ &- \lambda^2 \frac{(n+\nu)(2\alpha-\nu-1)}{y^2} \frac{A_n}{A_{n-1}} \xi_{n-1} \end{aligned} \tag{2.13}$$

Therefore, the action of the differential wave operator of Eq. (2.4c) on the basis element (2.12) gives the following

$$\begin{aligned}(H-E)\xi_n = \frac{\lambda^2}{2} &\left[ \frac{n}{y}\left(1 + \frac{\nu+1-2\alpha}{y}\right) + \frac{2E_\theta - \alpha(\alpha-1)}{y^2} + \frac{\alpha}{y} - \frac{1}{4} + \frac{2}{\lambda^2}(V_r - E) \right] \xi_n \\ &+ \frac{\lambda^2}{2} \frac{(n+\nu)(2\alpha-\nu-1)}{y^2} \frac{A_n}{A_{n-1}} \xi_{n-1} . \end{aligned} \tag{2.14}$$

The recurrence relation (A.6) and orthogonality relation (A.10) for the Laguerre polynomials show that a tridiagonal matrix representation $\langle \xi_n | H - E | \xi_{n'} \rangle$ is possible only for a limited number of special radial potential components $V_r$ and results in the following two possibilities:

(1) $\nu = 2\alpha - 1$, $\alpha(\alpha-1) = 2E_\theta$, and $V_r = \frac{\mathcal{Z}}{r}$ \hfill (2.15a)

(2) $\nu = 2\alpha - 2$, $\lambda^2 = -8E$, and $V_r = \frac{\mathcal{Z}}{r} + \frac{\mathcal{B}/2}{r^2}$ \hfill (2.15b)

where $\mathcal{Z}$ and $\mathcal{B}$ are real potential parameters: $\mathcal{Z}$ is the particle's charge and $\mathcal{B}$ is a centripetal potential barrier parameter. In what follows, we only consider the first case since the second is restricted (for real representations) to negative energies only. On the other hand, by taking the configuration coordinate in (2.12) as $y = (\lambda r)^2$, the basis function becomes compatible with the problem whose radial potential component is that of the oscillator, $V_r = \frac{1}{2}\omega^4 r^2$. This will be discussed briefly in Sec. 7.

## 3. Solution space for the angular component

### 3.1 Case (2.11a):

We start by studying the case (2.11a) for which the angular component of the non-central potential could be written as

$$V_\theta = \frac{\hat{C} + C\cos\theta}{2\sin^2\theta} - \frac{C_0}{2}\cos\theta. \qquad (3.1)$$

Thus the basis parameters $\mu$ and $\nu$ are related to the potential parameters $\hat{C}$ and $C$ and to the wavefunction quantum phase number $m$ by

$$\mu_m = \sqrt{m^2 + \hat{C} + C}, \quad \nu_m = \sqrt{m^2 + \hat{C} - C}. \qquad (3.2)$$

These relations show that $\mu$ and $\nu$ are discrete (indexed by $m$) and positive. Moreover, they also require that, for real representations, $m^2 \geq \max(-\hat{C} \pm C)$. That is, if we define $M$ as the smallest integer greater than $\max(|\hat{C} \pm C|)^{1/2}$, then

$$|m| = \begin{cases} 0, 1, 2, \ldots & , \max(\hat{C} \pm C) \geq 0 \\ M, M+1, M+2, \ldots & , \max(\hat{C} \pm C) < 0 \end{cases} \qquad (3.3)$$

It is worthwhile noting that a unique solution might exist for $m = \pm(M-1)$ in which the parameter(s) $\mu_m$ or (and) $\nu_m$ is (are) the negative of that (those) given by Eq. (3.2), and in the range $(-1, 0)$. This happens for special values of the potential parameters $\hat{C}$ and $C$ satisfying any (both) of the following two inequalities

$$1 - \hat{C} \pm C > (M-1)^2 > -\hat{C} \pm C. \qquad (3.4)$$

Now, to obtain the tridiagonal matrix representation $\langle \chi_n | H_\theta - E_\theta | \chi_{n'} \rangle$ we employ the orthogonality property (A.5) and recurrence relation (A.1) for the Jacobi polynomials into the action of the differential operator as given by Eq. (2.10). The result is as follows:

$$\langle \chi_n | H_\theta - E_\theta | \chi_{n'} \rangle = \left[ \frac{(\nu^2 - \mu^2)C_0/2}{(2n+\mu+\nu)(2n+\mu+\nu+2)} + \frac{1}{2}\left(n + \frac{\mu+\nu+1}{2}\right)^2 - \frac{1}{2}\left(\gamma + \frac{1}{2}\right)^2 \right] \delta_{n,n'}$$

$$+ \frac{C_0}{2n+\mu+\nu} \sqrt{\frac{n(n+\mu)(n+\nu)(n+\mu+\nu)}{(2n+\mu+\nu-1)(2n+\mu+\nu+1)}} \delta_{n,n'+1} \qquad (3.5)$$

$$+ \frac{C_0}{2n+\mu+\nu+2} \sqrt{\frac{(n+1)(n+\mu+1)(n+\nu+1)(n+\mu+\nu+1)}{(2n+\mu+\nu+1)(2n+\mu+\nu+3)}} \delta_{n,n'-1}$$

where we have introduced the dimensionless real parameter $\gamma$ by writing $2E_\theta \equiv \gamma(\gamma+1)$. The subscript $m$ on $\mu$ and $\nu$ was removed for the sake of clarity and simplicity in presentation. For arbitrary values of $C_0$, Eq. (3.5) shows that as the integers $n$ and $m$ increase the representation degenerates by changing its signature (becoming indefinite) when crossing the threshold defined by

$$\left(n + \frac{\mu_m + \nu_m + 1}{2}\right)^2 \leq \left(\gamma + \frac{1}{2}\right)^2. \qquad (3.6)$$

To keep the representation, which is bounded from below, definite and prevent it from degenerating we require that the set of integers $\{n \in \mathbb{N}, m \in \mathbb{Z}\}$ satisfy the constraint (3.6). Therefore, for a given $\gamma$, $m = 0, \pm 1, \pm 2, \ldots, \pm j$ [or $m = \pm M, \pm(M+1), \pm(M+2), \ldots, \pm j$, see Eq. (3.3)] such that $j$ is the maximum integer satisfying $\mu_j + \nu_j \leq |2\gamma+1| - 1$. Moreover, $n = 0, 1, 2, \ldots, N$, where $N$ is the maximum integer satisfying $N \leq \left|\gamma + \frac{1}{2}\right| -$

$\frac{\mu_0+\nu_0+1}{2}$ [or, $N \leq \left|\gamma+\frac{1}{2}\right| - \frac{\mu_M+\nu_M+1}{2}$]. Expanding the angular component of the wave function as $\Theta(\theta) = \sum_n f_n(E_\theta)\chi_n(x;\mu,\nu)$, then Eq. (2.4b) and Eq. (3.5) give the following three-term recursion relation for the expansion coefficients

$$z f_n = \left[\frac{\nu^2-\mu^2}{(2n+\mu+\nu)(2n+\mu+\nu+2)} + \frac{1}{C_0}\left(n+\frac{\mu+\nu+1}{2}\right)^2\right]f_n$$
$$+ \frac{2}{2n+\mu+\nu}\sqrt{\frac{n(n+\mu)(n+\nu)(n+\mu+\nu)}{(2n+\mu+\nu-1)(2n+\mu+\nu+1)}} f_{n-1} \qquad (3.7)$$
$$+ \frac{2}{2n+\mu+\nu+2}\sqrt{\frac{(n+1)(n+\mu+1)(n+\nu+1)(n+\mu+\nu+1)}{(2n+\mu+\nu+1)(2n+\mu+\nu+3)}} f_{n+1}$$

where $z = \frac{1}{C_0}\left(\gamma+\frac{1}{2}\right)^2$ and $C_0 \neq 0$. This recursion relation could be rewritten in terms of a three-parameter polynomial defined as

$$H_n^\sigma(z;\mu,\nu) = \frac{1}{\sqrt{2n+\mu+\nu+1}}\sqrt{\frac{\Gamma(n+\mu+1)\Gamma(n+\nu+1)}{\Gamma(n+1)\Gamma(n+\mu+\nu+1)}} f_n(z), \qquad (3.8)$$

where $\sigma = 1/C_0$. In terms of these orthogonal polynomials the recursion relation (3.7) becomes

$$z H_n^\sigma = \left[\frac{\nu^2-\mu^2}{(2n+\mu+\nu)(2n+\mu+\nu+2)} + \sigma\left(n+\frac{\mu+\nu+1}{2}\right)^2\right]H_n^\sigma$$
$$+ \frac{2(n+\mu)(n+\nu)}{(2n+\mu+\nu)(2n+\mu+\nu+1)} H_{n-1}^\sigma + \frac{2(n+1)(n+\mu+\nu+1)}{(2n+\mu+\nu+1)(2n+\mu+\nu+2)} H_{n+1}^\sigma \qquad (3.9)$$

These polynomials, as far as we know, have not been studied before. However, comparing this recurrence relation with (A.1) for the Jacobi polynomials, we can write

$$P_n^{(\mu,\nu)}(z) = \lim_{\sigma \to 0} H_n^\sigma(z;\mu,\nu). \qquad (3.10)$$

In this limit, $E_\theta$ should be allowed to increase such that the ratio $\left(2E_\theta + \frac{1}{4}\right)/C_0$, which is equal to $z$, becomes finite and belongs to the interval $[-1,+1]$. The polynomials $H_n^\sigma(z;\mu,\nu)$ are uniquely defined by the recursion (3.9) and the initial normalizing relation that $H_0^\sigma(z;\mu,\nu) = 1$. Now, the angular component of the wave function could then be written as the $L^2$–series sum

$$\Theta_m^a(\theta) = \mathcal{N}_m^a(z)\sum_{n=0}^{N_m} \frac{2n+\mu+\nu+1}{\sqrt{2^{\mu+\nu+1}}} \frac{\Gamma(n+1)\Gamma(n+\mu+\nu+1)}{\Gamma(n+\mu+1)\Gamma(n+\nu+1)} H_n^\sigma(z;\mu,\nu)(1-x)^{\frac{\mu}{2}}(1+x)^{\frac{\nu}{2}} P_n^{(\mu,\nu)}(x), (3.11a)$$

where $\mathcal{N}_m^a(z)$ is a normalization constant that depends on $z$, $m$ and the physical parameters of the problem but, otherwise, independent of $n$. To make $\Theta_m^a(\theta)$ $z$-normalizable, we write $\mathcal{N}_m^a(z) = \sqrt{\rho^\sigma(z)}$, where $\rho^\sigma(z)$ is the weight (density) function associated with the polynomials $\{H_n^\sigma\}$:

$$\int \rho^\sigma(z) H_n^\sigma(z;\mu,\nu) H_m^\sigma(z;\mu,\nu) dz = \frac{1}{2n+\mu+\nu} \frac{\Gamma(n+\mu+1)\Gamma(n+\nu+1)}{\Gamma(n+1)\Gamma(n+\mu+\nu+1)} \delta_{nm}. \qquad (3.12)$$

In Appendix B we show how to obtain this density function from the resolvent operator (Green's function) associated with these polynomials using the coefficients of the recursion relation, $\{a_n, b_n\}_{n=0}^\infty$. The sum in the series (3.11a) runs from $n = 0$ to $n = N_m$ where $N_m$, for a given $m$, is obtained from condition (3.6) as the largest integer $n$ satisfying the inequality

$$n \le \left|\gamma + \tfrac{1}{2}\right| - \tfrac{\mu_m + \nu_m + 1}{2}. \tag{3.6'}$$

The subscript $m$ appearing in $\Theta_m^a(\theta)$ is due to the fact that the parameters $\mu$ and $\nu$ depend on $m$ as shown in Eq. (3.2).

To obtain the diagonal representation,
$$\Theta_{nm}^a(\theta) = \chi_n(x; \mu_m, \nu_m) = \sqrt{\tfrac{2n+\mu_m+\nu_m+1}{2^{\mu_m+\nu_m+1}} \tfrac{\Gamma(n+1)\Gamma(n+\mu_m+\nu_m+1)}{\Gamma(n+\mu_m+1)\Gamma(n+\nu_m+1)}} (1-x)^{\tfrac{\mu_m}{2}} (1+x)^{\tfrac{\nu_m}{2}} P_n^{(\mu_m,\nu_m)}(x) \tag{3.13}$$
we impose the constraint (1.4) on Eq. (3.5) which gives
$$C_0 = 0, \quad \gamma_{nm}^\pm = -\tfrac{1}{2} \pm \left(n + \tfrac{\mu_m+\nu_m+1}{2}\right) = \begin{cases} n + \tfrac{\mu_m+\nu_m}{2} = \gamma \\ -n - \tfrac{\mu_m+\nu_m}{2} - 1 = -\gamma - 1 \end{cases} \tag{3.14}$$

According to Eq. (3.2), which states that $\mu_m, \nu_m > 0$, then $\gamma_{nm}^+ > 0$ and $\gamma_{nm}^- < -1$. Therefore, $\gamma > 0$ and $E_\theta > 0$ since $2E_\theta = \gamma(\gamma+1)$. Moreover, for a given value of $\gamma$ (equivalently, $E_\theta$), the above equation dictates that the integer $m$ must belong to the set $0, \pm 1, \pm 2, ..., \pm j$ [or the set $\pm M, \pm(M+1), \pm(M+2), ..., \pm j$, see Eq. (3.3)] where the integer $j$ satisfies the relation $\mu_j + \nu_j = 2\gamma$. This statement is the analogue of that, for central potentials, which says that $m = 0, \pm 1, \pm 2, ..., \pm \ell$, where $\ell$ is the orbital angular momentum quantum number. Additionally, for a given $m$ the integer $n$ should satisfy Eq. (3.14); that is, $n = \gamma - \tfrac{\mu_m+\nu_m}{2}$. This is also analogous to the central potential case where the principal quantum number is required to be equal to $\ell - |m|$, [6].

It is important to note that this diagonal representation is associated with the operator $H_\theta$. That is, $(H_\theta)_{nn'} = E_\theta \delta_{nn'}$. One should not confuse this with the discrete bound states spectrum, which is associated with the diagonal representation of the total Hamiltonian $H$ (i.e., $H_{nn'} = E \delta_{nn'}$). Consequently, it is not necessary to impose the constraints (3.14) on the bound states solution. That is, for bound states it is neither required that $C_0$ vanishes nor that $\gamma$ (equivalently, $E_\theta$) be quantized as $\gamma = (\mu_j + \nu_j)/2$. These points will be reemphasized when we construct the complete solution space in Sec. 5 and Sec. 6 below.

**3.2 Case (2.11b):**

The second case (2.11b) is associated with the following angular component of the non-central potential
$$V_\theta = \frac{\hat{C} + C \cos\theta}{2\sin^2\theta}, \tag{3.15}$$
where the real parameters $\mu$ and $C_1$ in (2.11b) are discretized via their relation to the integer $m$ and the potential parameters as follows
$$\mu_m = \sqrt{m^2 + \hat{C} + C}, \quad C_1 = m^2 + \hat{C} - C, \tag{3.16}$$
which requires that, for real representations, $m^2 \ge -\hat{C} - C$. Thus, if we define $M$ as the smallest integer greater than $\sqrt{|\hat{C}+C|}$, then
$$m = \begin{cases} 0, \pm 1, \pm 2, .. & , \hat{C}+C \ge 0 \\ \pm M, \pm(M+1), \pm(M+2), .. & , \hat{C}+C < 0 \end{cases} \tag{3.17}$$

It is also interesting to note that in this case, as well, a unique solution might exist for $m = \pm(M-1)$ in which the parameter $\mu_m$ is the negative of that given by Eq. (3.16) and its value will be in the range $(-1,0)$. This happens for special values of the potential parameters $\hat{C}$ and $C$ satisfying the following inequality

$$1 - \hat{C} - C > (M-1)^2 > -\hat{C} - C. \tag{3.18}$$

However, the dimensionless parameter $v$, aside from being larger than $-1$, is independent of $m$ and still arbitrary. The tridiagonal matrix representation $\langle \chi_n | H_\theta - E_\theta | \chi_{n'} \rangle$ for this case is

$$\langle \chi_n | H_\theta - E_\theta | \chi_{n'} \rangle =$$
$$\left\{ \frac{C_1}{4} - \left(\frac{v+1}{2}\right)^2 - \frac{n(n+\mu)}{2n+\mu+v} + \frac{2n(n+\mu+v+1)+(\mu+v)(v+1)}{(2n+\mu+v)(2n+\mu+v+2)} \left[ \left(n + \frac{\mu+v}{2} + 1\right)^2 - \tau^2 \right] \right\} \delta_{n,n'}$$
$$+ \frac{1}{2n+\mu+v} \sqrt{\frac{n(n+\mu)(n+v)(n+\mu+v)}{(2n+\mu+v-1)(2n+\mu+v+1)}} \left[ \left(n + \frac{\mu+v}{2}\right)^2 - \tau^2 \right] \delta_{n,n'+1} \tag{3.19}$$
$$+ \frac{1}{2n+\mu+v+2} \sqrt{\frac{(n+1)(n+\mu+1)(n+v+1)(n+\mu+v+1)}{(2n+\mu+v+1)(2n+\mu+v+3)}} \left[ \left(n + \frac{\mu+v}{2} + 1\right)^2 - \tau^2 \right] \delta_{n,n'-1}$$

where $\tau^2 = 2E_\theta + 1/4 = \left(\gamma + \frac{1}{2}\right)^2$ and, again, we took $2E_\theta = \gamma(\gamma+1)$. The subscript $m$ on $\mu$ was removed here, as well, for the sake of clarity in presentation. Here again, to prevent the representation from degenerating and becoming indefinite, we require that the set of integers $n$ and $m$ satisfy the constraint $\left(n + \frac{\mu_m + v}{2} + 1\right)^2 \leq \tau^2$. That is,

$$n + \frac{\mu_m + v}{2} + 1 \leq \left|\gamma + \frac{1}{2}\right|. \tag{3.20}$$

Therefore, for a given $\gamma$ and $v$, $m = 0, \pm 1, \pm 2, ..., \pm j$ [or $m = \pm M, \pm(M+1), \pm(M+2), ..., \pm j$, see Eq. (3.17)], where $j$ is the maximum integer that satisfies $\mu_j \leq |2\gamma+1| - v - 2$. Moreover, $n = 0, 1, 2, ..., N$, where $N$ is the maximum integer satisfying $N \leq \left|\gamma + \frac{1}{2}\right| - \frac{\mu_0 + v}{2} - 1$ [or, $N \leq \left|\gamma + \frac{1}{2}\right| - \frac{\mu_M + v}{2} - 1$]. Expanding the angular component of the wave function as $\Theta(\theta) = \sum_n f_n(E_\theta) \chi_n(x;\mu,v)$ and defining the three-parameter orthogonal polynomials

$$Q_n^\tau(z;\mu,v) = \frac{1}{\sqrt{2n+\mu+v+1}} \sqrt{\frac{\Gamma(n+\mu+1)\Gamma(n+v+1)}{\Gamma(n+1)\Gamma(n+\mu+v+1)}} f_n(z), \tag{3.21}$$

we obtain from Eq. (2.4b) and Eq. (3.19) the following three-term recursion relation that defines these polynomials

$$z Q_n^\tau = \left\{ -\frac{n(n+\mu)}{2n+\mu+v} + \frac{2n(n+\mu+v+1)+(\mu+v)(v+1)}{(2n+\mu+v)(2n+\mu+v+2)} \left[ \left(n + \frac{\mu+v}{2} + 1\right)^2 - \tau^2 \right] \right\} Q_n^\tau$$
$$+ \frac{(n+\mu)(n+v)}{(2n+\mu+v)(2n+\mu+v+1)} \left[ \left(n + \frac{\mu+v}{2}\right)^2 - \tau^2 \right] Q_{n-1}^\tau \tag{3.22}$$
$$+ \frac{(n+1)(n+\mu+v+1)}{(2n+\mu+v+1)(2n+\mu+v+2)} \left[ \left(n + \frac{\mu+v}{2} + 1\right)^2 - \tau^2 \right] Q_{n+1}^\tau$$

where $z = \left(\frac{v+1}{2}\right)^2 - \frac{C_1}{4}$. These polynomials are also new. However, comparing this recurrence relation with (A.1) of the Jacobi polynomials, we conclude that

$$P_n^{(\mu,v)}(z) = \lim_{\tau \to \infty} Q_n^\tau\left(-\tau^2 \frac{1+z}{2}; \mu, v\right). \tag{3.23}$$

The polynomials $Q_n^\tau(z;\mu,\nu)$ are uniquely defined by the recursion (3.22) and the initial normalizing relation that $Q_0^\tau(z;\mu,\nu) = 1$. Figure (1) gives a graphical illustration of the density function $\rho^\tau(z)$ associated with these polynomials for a given set of values of the parameters $\mu$, $\nu$, and $\tau$. The angular component of the wavefunction could then be written as the $L^2$–series sum

$$\Theta_m^b(\theta) = \mathcal{N}_m^b(z) \sum_{n=0}^{N_m} \frac{2n+\mu+\nu+1}{\sqrt{2^{\mu+\nu+1}}} \frac{\Gamma(n+1)\Gamma(n+\mu+\nu+1)}{\Gamma(n+\mu+1)\Gamma(n+\nu+1)} Q_n^\tau(z;\mu,\nu)(1-x)^{\frac{\mu}{2}}(1+x)^{\frac{\nu+1}{2}} P_n^{(\mu,\nu)}(x), (3.11b)$$

where $\mathcal{N}_m^b(z)$ is a normalization constant that depends on $z$, $m$ and the physical parameters. Normalizability of $\Theta_m^b(\theta)$ in $z$ space is achieved by taking $\mathcal{N}_m^b(z) = \sqrt{\rho^\tau(z)}$. For a given $m$, the integer index $N_m$ of the last term in the series (3.11b) is equal to the largest integer $n$ satisfying the constraint (3.20). Again, the presence of the subscript $m$ on $\Theta_m^b(\theta)$ is due to the dependence of $\mu$ and $z$ on $m$ as given by Eq. (3.16).

The diagonal representation is similarly obtained by imposing the constraint (1.4) on Eq. (3.19), which gives

$$\nu = -1 + \sqrt{m^2 + \hat{C} - C} = -1 + \nu_m, \quad \gamma_{nm}^\pm = -\tfrac{1}{2} \pm \left(n + \tfrac{\mu_m + \nu_m + 1}{2}\right), \qquad (3.24)$$

where $\nu_m$ is given by Eq. (3.2). Consequently, this diagonal representation is equivalent to the case (2.11a) above with $C_0 = 0$ and which is depicted by Eq. (3.14) and Eq. (3.13).

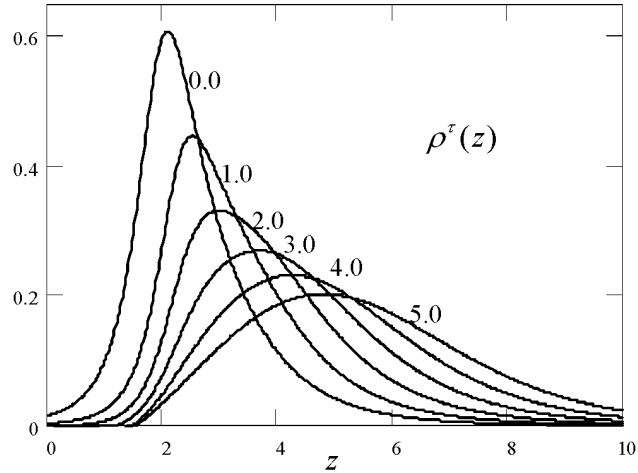

**Fig. 1**: The density (weight) function $\rho^\tau(z)$ associated with the orthogonal polynomials $Q_n^\tau(z;\mu,\nu)$. The "Dispersion Correction" method developed in [11] was used to generate this plot using the recursion coefficients $\{a_n,b_n\}_{n=0}^{50}$ obtained from Eq. (3.19) with $\mu = 1.0$, $\nu = 1.5$ and for several values of the parameter $\tau$ as shown on the traces.

### 3.3 Case (2.11c):

Analysis of this solution space, which corresponds to the case (2.11c), shows that it is identical to the second case (2.11b) if we make the following replacements:

$$C_1 \to C_2, \quad \mu \leftrightarrow \nu, \text{ and } x \to -x. \qquad (3.25)$$

Consequently, we can directly write the $L^2$–series solution for this case as follows

$$\Theta_m^c(\theta) = \mathcal{N}_m^c(z) \sum_{n=0}^{N_m} (-)^n \frac{2n+\mu+\nu+1}{\sqrt{2^{\mu+\nu+1}}} \frac{\Gamma(n+1)\Gamma(n+\mu+\nu+1)}{\Gamma(n+\mu+1)\Gamma(n+\nu+1)} Q_n^\tau(\tilde{z};\nu,\mu)(1-x)^{\frac{\mu+1}{2}}(1+x)^{\frac{\nu}{2}} P_n^{(\mu,\nu)}(x), \quad (3.11c)$$

where $\tilde{z} = \left(\frac{\mu+1}{2}\right)^2 - \frac{C_2}{4}$. The normalization constant $\mathcal{N}_m^c$ is obtained from $\mathcal{N}_m^b$ by the prescription (3.25) and we have used the Jacobi polynomials identity: $P_n^{(\mu,\nu)}(-x) = (-)^n \times P_n^{(\nu,\mu)}(x)$. The presence of the subscript $m$ on $\Theta_m^c(\theta)$ is due to the dependence of $\nu$ and $\tilde{z}$ on $m$. The diagonal representation is obtained for the parameter values given by Eq. (3.24) with the exchange $\mu \leftrightarrow \nu$.

## 4. Solution space for the radial component

To obtain the total wavefunction we only need to calculate the remaining radial component $R(r)$ or, equivalently, its expansion coefficients $\{g_n\}_{n=0}^\infty$. To that end, we expand $|R\rangle$ in $(H-E)|R\rangle = 0$ as $\sum_m g_m |\xi_m\rangle$ and project on the left by $\langle \xi_n |$. This results in a three-term recursion relation that will be solved for $g_n$ in terms of orthogonal polynomials. We only consider the situation described by (2.15a) which applies to the Coulomb potential and to all energies, positive and negative. If we rewrite the real dimensionless angular separation parameter $E_\theta$ as $2E_\theta = \gamma(\gamma+1)$, then the radial wave equation (2.4c) could be recast as follows

$$\left[\frac{d^2}{dr^2} - \frac{\gamma(\gamma+1)}{r^2} - 2V_r + 2E\right] R = 0. \quad (4.1)$$

Hence, the parameter $\gamma$ plays the role of the orbital angular momentum in problems with central (spherically symmetric) potentials. However, writing $2E_\theta$ as $\gamma(\gamma+1)$ implies that, for real values, $E_\theta$ is restricted to the range $E_\theta \geq -1/8$. Now, for positive values of $E_\theta$, we either have $\gamma > 0$, which we refer to as $\gamma^+$, or $\gamma < -1$, which we call $\gamma^-$. For the case where the representation of the angular component is tridiagonal, $\gamma^\pm$ is a continuous parameter. However, for the diagonal angular representation it assumes discrete values indexed by $n$ and $m$ as given by Eq. (3.14). We start our investigation with the general continuous angular parameter $\gamma$. Consequently, the parameter $\alpha$ for the regular solution of the radial wave function could then be written in terms of $\gamma^\pm$ using (2.15a) as follows

$$\alpha = \begin{cases} \gamma^+ + 1 & , \gamma > 0 \\ -\gamma^- & , \gamma < -1 \end{cases} \quad (4.2)$$

which implies that $\alpha$ is always greater than +1. Now, to obtain the sought after recursion relation for the expansion coefficients of the radial wavefunction, we utilize the action of the wave operator on the basis as given by Eq. (2.14) and the parameter assignments (2.15a). This results (with the use of the orthogonality and recurrence relations of the Laguerre polynomials) in the following tridiagonal matrix representation

$$\frac{2}{\lambda^2} \langle \xi_k | H - E | \xi_{k'} \rangle = \left[2(k+\alpha)\left(\frac{1}{4} - \frac{2E}{\lambda^2}\right) + \frac{2Z}{\lambda}\right] \delta_{k,k'} + \left(\frac{1}{4} + \frac{2E}{\lambda^2}\right)\left[\sqrt{k(k+2\alpha-1)}\delta_{k,k'+1} + \sqrt{(k+1)(k+2\alpha)}\delta_{k,k'-1}\right], \quad (4.3)$$

where $k, k' = 0, 1, 2, \ldots$. Therefore, the resulting recursion relation (1.3) for the expansion coefficients of the radial wavefunction becomes

$$\left[ 2(k+\alpha) \frac{\sigma_-}{\sigma_+} - \frac{2\mathcal{Z}/\lambda}{\sigma_+} \right] g_k - \sqrt{k(k+2\alpha-1)}\, g_{k-1} - \sqrt{(k+1)(k+2\alpha)}\, g_{k+1} = 0, \quad (4.4)$$

where $\sigma_\pm = \frac{2E}{\lambda^2} \pm \frac{1}{4}$. Rewriting this recursion in terms of the polynomials $P_k(E) = \sqrt{\Gamma(k+2\alpha)/\Gamma(k+1)}\, g_k(E)$, we obtain a more familiar recursion relation as follows

$$2\left[ (k+\alpha) \frac{\sigma_-}{\sigma_+} - \frac{\mathcal{Z}/\lambda}{\sigma_+} \right] P_k - (k+2\alpha-1) P_{k-1} - (k+1) P_{k+1} = 0. \quad (4.5)$$

We compare this with the three-term recursion relation for the Meixner-Pollaczek polynomials $P_k^\mu(z, \varphi)$ [7]:

$$2\left[ (k+\mu)\cos\varphi + z\sin\varphi \right] P_k^\mu - (k+2\mu-1) P_{k-1}^\mu - (k+1) P_{k+1}^\mu = 0, \quad (4.6)$$

where $\mu > 0$ and $0 < \varphi < \pi$. The comparison is valid only within this permissible range of values of the parameters. This means that the solution obtained as such is valid only for $E > 0$ (i.e., for the continuum scattering states). Thus for the continuum case, we obtain the following two alternative $L^2$–series solutions (depending on the value of the parameter $\gamma$) for the radial component of the wavefunction

$$R^{\gamma^+}(r) = \mathcal{N}^{\gamma^+}(E) \sum_{k=0}^\infty \frac{\Gamma(k+1)}{\Gamma(k+2\gamma^+ + 2)} P_k^{\gamma^+ + 1}\left( \frac{-\mathcal{Z}}{\sqrt{2E}}, \cos^{-1} \frac{E - \lambda^2/8}{E + \lambda^2/8} \right) (\lambda r)^{\gamma^+ + 1} e^{-\lambda r/2} L_k^{2\gamma^+ + 1}(\lambda r), \quad (4.7a)$$

$$R^{\gamma^-}(r) = \mathcal{N}^{\gamma^-}(E) \sum_{k=0}^\infty \frac{\Gamma(k+1)}{\Gamma(k - 2\gamma^-)} P_k^{-\gamma^-}\left( \frac{-\mathcal{Z}}{\sqrt{2E}}, \cos^{-1} \frac{E - \lambda^2/8}{E + \lambda^2/8} \right) (\lambda r)^{-\gamma^-} e^{-\lambda r/2} L_k^{-2\gamma^- - 1}(\lambda r), \quad (4.7b)$$

where $\mathcal{N}^{\gamma^\pm}$ are normalization constants that depend only on the physical parameters of the problem. To make $R^{\gamma^\pm}(r)$ energy-normalizable we write $\mathcal{N}^\gamma = \sqrt{\lambda \mathcal{Z} \rho^\gamma(z)/(2E)^{3/2}}$, where $\rho^\gamma(z)$ is the density function associated with the orthogonality of the Meixner-Pollaczek polynomials:

$$\int \rho^\mu(z) P_n^\mu(z, \varphi) P_m^\mu(z, \varphi)\, dz = \frac{\Gamma(n+2\mu)}{\Gamma(n+1)} \delta_{nm}. \quad (4.8)$$

On the other hand, imposing the diagonalization constraints (1.4) on the tridiagonal matrix representation (4.3) gives the following energy spectrum

$$E_k = -\mathcal{Z}^2 / (k+\alpha)^2 = \begin{cases} -\mathcal{Z}^2 / (k+\gamma^+ + 1)^2 \\ -\mathcal{Z}^2 / (k-\gamma^-)^2 \end{cases} \quad (4.9)$$

and requires that the length scale parameter $\lambda$ be discretized as follows

$$\lambda_k = -2\mathcal{Z} / (k+\alpha) = \begin{cases} -2\mathcal{Z} / (k+\gamma^+ + 1) \\ -2\mathcal{Z} / (k-\gamma^-) \end{cases} \quad (4.10)$$

which requires that $\mathcal{Z} < 0$ since $\lambda$ must be positive (i.e., bound states exist only for an attractive coulomb potential). The corresponding radial component of the discrete bound states wavefunction is

$$R_k^+(r) = \xi_k(y; \lambda_k^+, \nu^+) = \sqrt{\frac{\lambda_k^+ \Gamma(k+1)}{\Gamma(k+2\gamma^+ + 2)}} (\lambda_k^+ r)^{\gamma^+ + 1} e^{-\lambda_k^+ r/2} L_k^{2\gamma^+ + 1}(\lambda_k^+ r), \quad (4.11a)$$

$$R_k^-(r) = \xi_k(y; \lambda_k^-, \nu^-) = \sqrt{\frac{\lambda_k^- \Gamma(k+1)}{\Gamma(k-2\gamma^-)}} (\lambda_k^- r)^{-\gamma^-} e^{-\lambda_k^- r/2} L_k^{-2\gamma^- - 1}(\lambda_k^- r), \quad (4.11b)$$

where $\lambda_k^\pm$ is given by Eq. (4.10) above.

Now, if the angular representation is diagonal, then $\gamma^{\pm}$ assumes the discrete values given by Eq. (3.14). In this case, the radial component of the wavefunction for the continuum scattering states could be written as

$$R_{nm}(r) = \mathcal{N}_{nm}(E) \sum_{k=0}^{\infty} \left[ \frac{\Gamma(k+1)}{\Gamma(k+2n+\mu_m+\nu_m+2)} P_k^{n+\frac{\mu_m+\nu_m}{2}+1}\left(\frac{-\mathcal{Z}}{\sqrt{2E}}, \cos^{-1}\frac{E-\lambda^2/8}{E+\lambda^2/8}\right) \right.$$
$$\left. \times (\lambda r)^{n+\frac{\mu_m+\nu_m}{2}+1} e^{-\lambda r/2} L_k^{2n+\mu_m+\nu_m+1}(\lambda r) \right] \quad (4.12)$$

Whereas, for bound states, it reads

$$R_{knm}(r) = \xi_k(y; \lambda_{knm}, \nu_{knm})$$
$$= \sqrt{\frac{\lambda_{knm} \Gamma(k+1)}{\Gamma(k+2n+\mu_m+\nu_m+2)}} (\lambda_{knm} r)^{n+\frac{\mu_m+\nu_m}{2}+1} e^{-\lambda_{knm} r/2} L_k^{2n+\mu_m+\nu_m+1}(\lambda_{knm} r) \quad (4.13)$$

where $\lambda_{knm} = -2\mathcal{Z}/\left(k+n+\frac{\mu_m+\nu_m}{2}+1\right)$ and the corresponding energy spectrum is

$$E_{knm} = -\mathcal{Z}^2 / \left(k+n+\frac{\mu_m+\nu_m}{2}+1\right)^2. \quad (4.14)$$

In the following two sections we use the above findings to construct the complete solution space for the continuum scattering states and for the discrete bound states. Two alternative solutions are obtained depending on whether or not the potential parameter $C_0$ vanishes.

## 5. The complete solution space for $C_0 \neq 0$

For the continuum scattering states, this solution space is parameterized by three quantum numbers and six real parameters. The real parameters are $\{\hat{C}, C, C_0, \mathcal{Z}, \gamma, E\}$, where $E > 0$ and either $\gamma = \gamma^+ > 0$ or $\gamma = \gamma^- < -1$. The quantum numbers are $\{k, n, m\}$, where $k \in \mathbb{N} = 0, 1, 2, ..,$ and $m \in \mathbb{Z}^j = 0, \pm 1, ..., \pm j$ [or $m = \pm M, \pm(M+1), \pm(M+2), ..., \pm j$, see Eq. (3.3)]. $j$ is the largest integer satisfying $\mu_j + \nu_j \leq 2\gamma^+$ or $\mu_j + \nu_j \leq -2(\gamma^- + 1)$ and $M$ is the smallest integer greater than $\max(|\hat{C} \pm C|)^{1/2}$. For a given integer $m$, $n = 0, 1, 2, ..., N_m$, where $N_m$ is the largest integer $n$ that satisfies either $n \leq \gamma^+ - \frac{\mu_m+\nu_m}{2}$ or $n \leq -\gamma^- - 1 - \frac{\mu_m+\nu_m}{2}$. It might be worthwhile to compare this with the spherically symmetric Coulomb problem [1,8] whose continuum solution space is parameterized by four quantum numbers $\{k, n, m, \ell\}$ and two real parameters $\{\mathcal{Z}, E\}$, where $E > 0$, $k, \ell \in \mathbb{N}$, and $m \in \mathbb{Z}^\ell = 0, \pm 1, ..., \pm \ell$ with $n = 0, 1, 2, ..., \ell - |m|$. Now, the total wave function, which is an element in this space, is written in terms of the radial component $R^{\gamma^\pm}(r)$ in (4.7ab) and the angular components, $\Theta_m^a(\theta)$ in (3.11a) and $\Phi_m(\phi)$ in (2.6).

For bound states, the energy spectrum is quantized via its dependence on a single quantum number as given by Eq. (4.9) and the complete bound states wavefunction is written in terms of the angular components, $\Phi_m(\phi)$ in (2.6) and $\Theta_m^a(\theta)$ in (3.11a), and the radial component in (4.11ab).

It is worthwhile to consider the special case where $\hat{C} = C = 0$ and $C_0 \neq 0$ which is unique to our work. In this case, Eq. (3.2) gives $\mu_m = \nu_m = |m|$. If we write the positive (negative) parameter $\gamma$ as $\gamma^+ = j + \eta$ ($\gamma^- = -j - 1 - \eta$), where $0 \leq \eta < 1$, then the complete scattering state wave function becomes

$$\psi_j(\vec{r}, E) = \mathcal{N} \sum_{k=0}^{\infty} \sum_{n=0}^{j} \sum_{m=n-j}^{j-n} \left[ \frac{(n+|m|+1/2)\Gamma(k+1)}{2^{|m|}\Gamma(k+2j+2\eta+2)} \frac{\Gamma(n+1)\Gamma(n+2|m|+1)}{\Gamma(n+|m|+1)^2} P_k^{j+\eta+1}\left(\frac{-\mathcal{Z}}{\sqrt{2E}}, \cos^{-1}\frac{E - \lambda^2/8}{E + \lambda^2/8}\right) \right. \tag{5.1}$$
$$\left. \times H_n^{|m|}(z; C_0) \times (\lambda r)^{j+\eta+1} e^{-\lambda r/2} (1-x^2)^{\frac{|m|}{2}} \times L_k^{2j+2\eta+1}(\lambda r) \times P_n^{(|m|,|m|)}(x) \times e^{im\phi} \right]$$

where the normalization constant $\mathcal{N} = \sqrt{\lambda \mathcal{Z} \rho^\sigma(z)\rho^\gamma\left(\frac{-\mathcal{Z}}{\sqrt{2E}}\right)/\pi(2E)^{3/2}}$, $z = \frac{1}{C_0}\left(\eta + j + \frac{1}{2}\right)^2$ and the polynomials $H_n^{|m|}(z; C_0)$ satisfy the three-term recursion relation

$$z H_n^{|m|} = \frac{1}{C_0}\left(n + |m| + \frac{1}{2}\right)^2 H_n^{|m|} + \frac{n+|m|}{2(n+|m|+1/2)} H_{n-1}^{|m|} + \frac{(n+1)(n+2|m|+1)}{2(n+|m|+1/2)(n+|m|+1)} H_{n+1}^{|m|}. \tag{3.9'}$$

The bound states energy spectrum for this special case is $E_{kj} = -\mathcal{Z}^2/(k+j+\eta+1)^2$ and the corresponding wavefunction is

$$\psi_{kj}(\vec{r}) = \mathcal{N}_k \sum_{n=0}^{j} \sum_{m=n-j}^{j-n} \left[ \frac{(n+|m|+1/2)}{2^{|m|}} \frac{\Gamma(n+1)\Gamma(n+2|m|+1)}{\Gamma(n+|m|+1)^2} H_n^{|m|}(z; C_0) \right. \tag{5.2}$$
$$\left. \times (\lambda_{kj} r)^{j+\eta+1} e^{-\lambda_{kj} r/2} (1-x^2)^{\frac{|m|}{2}} \times L_k^{2j+2\eta+1}(\lambda_{kj} r) \times P_n^{(|m|,|m|)}(x) \times e^{im\phi} \right]$$

where $\lambda_{kj} = -2\mathcal{Z}/(k+j+\eta+1)$ and $\mathcal{N}_k = \sqrt{\lambda \rho^\sigma(z)\Gamma(k+1)/\pi\Gamma(k+2j+2\eta+2)}$.

## 6. The complete solution space for $C_0 = 0$

This space splits into four subspaces depending on the values of the potential parameters, $\hat{C}$ and $C$:

**6.1** $\hat{C} \pm C \geq 0$: The angular component of the solution space carries the diagonal representation $\Theta_{nm}^a(\theta) = \chi_n(x; \mu_m, \nu_m)$ given by Eq. (3.13), which is associated with the case (2.11a), where $m = 0, \pm 1, \pm 2, ..., \pm j$, $\mu_j + \nu_j = 2\gamma$ and, for a given $m$, $n = \gamma - \frac{\mu_m + \nu_m}{2}$. The other angular component of the wavefunction is $\Phi_m(\phi)$ which is given by (2.6). For the continuum scattering states, the radial component is given by Eq. (4.12), whereas, for the discrete bound states it is given by Eq. (4.13) and the corresponding energy spectrum is given by Eq. (4.14).

**6.2** $\hat{C} \pm C < 0$: This case is identical to the previous one except that the angular phase quantum number $m$ belongs to the range of values $m = \pm M, \pm(M+1), \pm(M+2), ..., \pm j$, where $M$ is the smallest integer that is greater than $\max(|\hat{C} \pm C|)^{1/2}$.

Almost all of the work that has been reported in the literature about exact solutions of this kind of non-central problems (with $V_\phi = 0$) is confined to either one of the above two representations, where the angular component is diagonal, $\alpha = \mu_m/2$, $\beta = \nu_m/2$, and $C_0 = 0$. Moreover, some of these solutions were obtained using path integral formulation utilizing the Kustannheimo-Stiefel transformation. Additionally, most are dealing with bound states while others were solved only in parabolic coordinates. For more recent examples of such work, one may consult the papers in [9] and references therein. The article by Khare and Bhaduri [10] is a good introduction to the general problem of non-central potentials in two and three dimensions.

**6.3** $C > \hat{C} \geq -C$: This solution space carries the tridiagonal representation $\Theta_m^b(\theta)$ given by Eq. (3.11b), which is associated with the case (2.11b), where $m = 0, \pm 1, \pm 2, ..., \pm j$ and $j$ is the maximum integer that satisfies $\mu_j \leq |2\gamma + 1| - \nu - 2$. Moreover, for a given $\gamma$ and $\nu$, $N_m$ is equal to the largest integer $n$ satisfying $n \leq |\gamma + \frac{1}{2}| - \frac{\mu_m + \nu}{2} - 1$. The other angular component $\Phi_m(\phi)$ is given by (2.6). The radial component for the continuum scattering states is given by Eq. (4.7ab), whereas, for the discrete bound states it is given by Eq. (4.11ab) and the corresponding energy spectrum is that given by Eq. (4.9).

**6.4** $-C > \hat{C} \geq C$: The angular component of this solution space is tridiagonal and corresponds to the case (2.11c). Consequently, it is identical to the previous case after the application of the map given by (3.25). For example, $j$ is the maximum integer that satisfies $\nu_j \leq |2\gamma + 1| - \mu - 2$ and $N_m$ is the largest integer $n$ satisfying $n \leq |\gamma + \frac{1}{2}| - \frac{\nu_m + \mu}{2} - 1$.

## 7. Solution for the radial oscillator potential

In this section we give a brief treatment of the same problem but with the radial component of the non-central potential (2.3) being the oscillator potential $V_r = \frac{1}{2}\omega^4 r^2$, where $\omega$ is the oscillator frequency. This is accomplished by taking the configuration space coordinate $y = (\lambda r)^2$ in the expression of the $L^2$ basis elements (2.12). Using the differential equation and differential formula for the Laguerre polynomials, the action of the wave operator of Eq. (2.4c) with $2E_\theta = \gamma(\gamma + 1)$ on this basis gives the following

$$(H - E)\xi_n = 2\lambda^2 \left[ -\frac{n(2\alpha - \nu - \frac{1}{2}) + (\alpha - \frac{1}{4})^2 - \frac{1}{4}(\gamma + \frac{1}{2})^2}{y} + n + \alpha + \frac{1}{4} - \frac{y}{4} \right.$$
$$\left. + \frac{1}{2\lambda^2}(V_r - E) \right] \xi_n + 2\lambda^2 \frac{(n+\nu)(2\alpha - \nu - \frac{1}{2})}{y} \frac{A_n}{A_{n-1}} \xi_{n-1} \,. \quad (7.1)$$

The recurrence relation and orthogonality property of the Laguerre polynomials show that a tridiagonal matrix representation $\langle \xi_n | H - E | \xi_{n'} \rangle$ is possible only for a limited number of special radial potential components $V_r$ and results in the following two possibilities:

(1) $\nu = 2\alpha - 1/2$, $2\alpha = \begin{cases} \gamma + 1 \\ -\gamma \end{cases}$, and $V_r = \frac{1}{2}\omega^4 r^2$ \hfill (7.2a)

(2) $\nu = 2\alpha - 3/2$, and $V_r = \frac{1}{2}\lambda^4 r^2 + \mathcal{B}/2r^2$ \hfill (7.2b)

where $\omega$ is the oscillator frequency and $\mathcal{B}$ is a centripetal potential barrier parameter. We only consider the first case (7.2a) which results in the following tridiagonal matrix representation of the wave operator

$$\frac{2}{\lambda^2}\langle\xi_k|H-E|\xi_{k'}\rangle = \left[(2k+\nu+1)\left(\frac{\omega^4}{\lambda^4}+1\right) - \frac{2E}{\lambda^2}\right]\delta_{k,k'}$$
$$-\left(\frac{\omega^4}{\lambda^4}-1\right)\left[\sqrt{k(k+\nu)}\delta_{k,k'+1} + \sqrt{(k+1)(k+\nu+1)}\delta_{k,k'-1}\right]. \quad (7.3)$$

This representation leads to a three-term recursion relation for the expansion coefficients of the radial component of the wavefunction which could be solved in terms of a "Hyperbolic-type" Meixner-Pollaczek polynomial defined as

$$\tilde{P}_k^\mu(z,\varphi) \equiv P_k^\mu(-iz,i\varphi) = \frac{\Gamma(k+2\mu)}{\Gamma(k+1)\Gamma(2\mu)} e^{-k\varphi} {}_2F_1(-k,\mu+z;2\mu;1-e^{2\varphi}), \quad (7.4)$$

where $\varphi > 0$ and ${}_2F_1$ is the hypergeometric function. These polynomials satisfy the following modified [with respect to the original relation (4.6) above] three-term recursion relation:

$$2\left[(k+\mu)\cosh\varphi + z\sinh\varphi\right]\tilde{P}_k^\mu - (k+2\mu-1)\tilde{P}_{k-1}^\mu - (k+1)\tilde{P}_{k+1}^\mu = 0. \quad (7.5)$$

However, we will not pursue this line of investigation and be contended with only obtaining the discrete bound state wavefunction and energy spectrum. This requires that the matrix representation of the Hamiltonian be diagonal, which when imposed on (7.3) gives $\lambda^2 = \omega^2$ and the following energy spectrum

$$E_k^\pm = \begin{cases} \omega^2(2k+\gamma^+ + 3/2) \\ \omega^2(2k-\gamma^- + 1/2) \end{cases} \quad (7.6)$$

The total bound states wavefunction for the case $C_0 \neq 0$ is written in terms of the angular components $\Phi_m(\phi)$ in (2.6), $\Theta_m^a(\theta)$ in (3.11a) and the following radial component

$$R_k^+(r) = \sqrt{\frac{2\omega\Gamma(k+1)}{\Gamma(k+\gamma^++3/2)}}(\omega r)^{\gamma^++1} e^{-\omega^2 r^2/2} L_k^{\gamma^++1/2}(\omega^2 r^2), \quad (7.7a)$$

$$R_k^-(r) = \sqrt{\frac{2\omega\Gamma(k+1)}{\Gamma(k-\gamma^-+1/2)}}(\omega r)^{-\gamma^-} e^{-\omega^2 r^2/2} L_k^{-\gamma^--1/2}(\omega^2 r^2). \quad (7.7b)$$

**Acknowledgments**

The author dedicates this work to Abu Khaled for the continued support. The help provided by M. S. Abdelmonem and F. A. Al-Sulaiman in literature search is gratefully acknowledged. Valuable and motivating discussions with H. A. Yamani are highly appreciated.**Appendix A: The Laguerre and Jacobi polynomials**

The following are useful formulas and relations associated with these polynomials. They are found in most books on orthogonal polynomials but listed here for ease of reference.

(1) The Jacobi polynomials $P_n^{(\mu,\nu)}(x)$, where $\mu > -1, \nu > -1$:

$$\left(\frac{1\pm x}{2}\right)P_n^{(\mu,\nu)} = \frac{2n(n+\mu+\nu+1)+(\mu+\nu)(\frac{\mu+\nu}{2}\pm\frac{\nu-\mu}{2}+1)}{(2n+\mu+\nu)(2n+\mu+\nu+2)}P_n^{(\mu,\nu)}$$
$$\pm\frac{(n+\mu)(n+\nu)}{(2n+\mu+\nu)(2n+\mu+\nu+1)}P_{n-1}^{(\mu,\nu)} \pm \frac{(n+1)(n+\mu+\nu+1)}{(2n+\mu+\nu+1)(2n+\mu+\nu+2)}P_{n+1}^{(\mu,\nu)} \quad (A.1)$$

$$P_n^{(\mu,\nu)}(x) = \frac{\Gamma(n+\mu+1)}{\Gamma(n+1)\Gamma(\mu+1)}\,{}_2F_1(-n, n+\mu+\nu+1; \mu+1; \tfrac{1-x}{2}) = (-)^n P_n^{(\nu,\mu)}(-x) \quad (A.2)$$

$$\left\{(1-x^2)\frac{d^2}{dx^2} - \left[(\mu+\nu+2)x+\mu-\nu\right]\frac{d}{dx} + n(n+\mu+\nu+1)\right\}P_n^{(\mu,\nu)}(x) = 0 \quad (A.3)$$

$$(1-x^2)\frac{d}{dx}P_n^{(\mu,\nu)} = -n\left(x+\frac{\nu-\mu}{2n+\mu+\nu}\right)P_n^{(\mu,\nu)} + 2\frac{(n+\mu)(n+\nu)}{2n+\mu+\nu}P_{n-1}^{(\mu,\nu)} \quad (A.4)$$

$$\int_{-1}^{+1}(1-x)^\mu(1+x)^\nu P_n^{(\mu,\nu)}(x)P_m^{(\mu,\nu)}(x)dx = \frac{2^{\mu+\nu+1}}{2n+\mu+\nu+1}\frac{\Gamma(n+\mu+1)\Gamma(n+\nu+1)}{\Gamma(n+1)\Gamma(n+\mu+\nu+1)}\delta_{nm} \quad (A.5)$$

(2) The Laguerre polynomials $L_n^\nu(x)$, where $\nu > -1$:

$$xL_n^\nu = (2n+\nu+1)L_n^\nu - (n+\nu)L_{n-1}^\nu - (n+1)L_{n+1}^\nu \quad (A.6)$$

$$L_n^\nu(x) = \frac{\Gamma(n+\nu+1)}{\Gamma(n+1)\Gamma(\nu+1)}\,{}_1F_1(-n;\nu+1;x) \quad (A.7)$$

$$\left[x\frac{d^2}{dx^2} + (\nu+1-x)\frac{d}{dx} + n\right]L_n^\nu(x) = 0 \quad (A.8)$$

$$x\frac{d}{dx}L_n^\nu = nL_n^\nu - (n+\nu)L_{n-1}^\nu \quad (A.9)$$

$$\int_0^\infty x^\nu e^{-x} L_n^\nu(x)L_m^\nu(x)dx = \frac{\Gamma(n+\nu+1)}{\Gamma(n+1)}\delta_{nm} \quad (A.10)$$

## Appendix B: The orthogonal polynomials $H_n^\sigma(z;\mu,\nu)$ and $Q_n^\tau(z;\mu,\nu)$

For the purpose of economy in presentation we make a simultaneous treatment of both polynomials by studying the expansion coefficients (polynomials) $\{f_n(z)\}_{n=0}^\infty$ which are related to these two polynomials by the definitions (3.8) and (3.21), respectively. The recursion relation (1.3) together with the initial seed

$$f_0 = \sqrt{\mu+\nu+1}\sqrt{\frac{\Gamma(\mu+\nu+1)}{\Gamma(\mu+1)\Gamma(\nu+1)}}, \quad (B.1)$$

which is obtained from the normalization $H_0^\sigma = Q_0^\tau = 1$, give a unique definition of these polynomials. In fact, one obtains recursively

$$\begin{aligned}
f_1(z) &= \tfrac{z-a_0}{b_0}f_0 \\
f_2(z) &= \tfrac{1}{b_1}\left[(z-a_1)f_1(z) - b_0 f_0\right] \\
f_3(z) &= \tfrac{1}{b_2}\left[(z-a_2)f_2(z) - b_1 f_1(z)\right]
\end{aligned} \quad (B.2)$$

....

These polynomials are orthonormal with respect to the measure $\rho(z)dz$, where $\rho(z)$ is the weight (density) function associated with these polynomials. That is,

$$\int_{z_-}^{z_+} \rho(z) f_n(z) f_m(z) dz = \delta_{nm}, \tag{B.3a}$$

where $z_\mp$ is the left/right boundary of the one-dimensional real space with coordinate $z$. This orthogonality relation could also be written in terms of the target polynomials (e.g., $H_n^\sigma$) as follows:

$$\int_{z_-}^{z_+} \rho^\sigma(z) H_n^\sigma(z;\mu,\nu) H_m^\sigma(z;\mu,\nu) dz = \frac{1}{2n+\mu+\nu} \frac{\Gamma(n+\mu+1)\Gamma(n+\nu+1)}{\Gamma(n+1)\Gamma(n+\mu+\nu+1)} \delta_{nm}. \tag{B.3b}$$

One way to obtain the density function is to use its relationship to the resolvent operator (Green's function), $G(z)$, associated with this system of polynomials:

$$\rho(z) = \frac{1}{2\pi i}\left[ G(z+i0^+) - G(z-i0^-) \right] = \frac{1}{\pi} \operatorname{Im} G(z+i0^+). \tag{B.4}$$

That is, the density is equal to the discontinuity of the Green's function across the cut on the real $z$-axis in the complex plane. One of the representations of the Green's function could be written in terms of the recursion coefficients $\{a_n, b_n\}_{n=0}^\infty$ as the infinite continued fraction:

$$G(z) = \cfrac{-1}{z - a_0 - \cfrac{b_0^2}{z - a_1 - \cfrac{b_1^2}{z - a_2 \dots}}} \tag{B.5}$$

Assuming that the system under study has a single density band with finite width, then there exist a single limit for the recursion coefficients:

$$a_\infty = \lim_{n\to\infty} a_n, \quad b_\infty = \lim_{n\to\infty} b_n. \tag{B.6}$$

In such a case the boundary of the band (the ends of the non-zero support interval in the real space $z$) is

$$z_\pm = a_\infty \pm 2b_\infty, \tag{B.7}$$

and the resolvent operator (B.5) could be approximated by

$$G(z) \approx \cfrac{-1}{z - a_0 - \cfrac{b_0^2}{z - a_1 - \cfrac{b_1^2}{z - a_2 \dots - \cfrac{b_{N-2}^2}{z - a_{N-1} - T(z)}}}} \tag{B.8}$$

for some large enough integer $N$ such that $|a_\infty - a_N| \approx 0$, $|b_\infty - b_{N-1}| \approx 0$. The "terminator" function $T(z)$ is given by

$$T(z) = \cfrac{b_\infty^2}{z - a_\infty - \cfrac{b_\infty^2}{z - a_\infty - \dots}} = \frac{b_\infty^2}{z - a_\infty - T(z)} \tag{B.9}$$

which could be solved for $T(z)$ as

$$T(z) = \frac{z - a_\infty}{2} \pm \frac{1}{2}\sqrt{(z - a_\infty - 2b_\infty)(z - a_\infty + 2b_\infty)}. \tag{B.10}$$

Therefore, it becomes obvious from this expression that $\operatorname{Im} T(z) = 0$ results in $\rho(z) = 0$ outside the density band and gives its boundaries as $a_\infty \pm 2b_\infty$. However, in the present case such limits do not exist. Specifically, for a large integer $N$ we obtain from Eq. (3.7) the following values associated with the polynomials $\{H_n^\sigma\}$

$$a_N \approx \sigma N^2,\ b_N \approx 1/2, \tag{B.11}$$

whereas, for the polynomials $\{Q_n^\tau\}$, these could be obtained from (3.19) as follows

$$a_N \approx N^2/2,\ b_N \approx N^2/4. \tag{B.12}$$

Consequently, we could infer that the interval on the real $z$-axis associated with the polynomials $\{Q_n^\tau\}$, is the semi-infinite $z \in [0,\infty]$, whereas for $\{H_n^\sigma\}$ the corresponding interval is $\sigma$–dependent (for $\sigma \to 0$, $z \in [-1,+1]$). These results are also supported by Eq. (3.23) and Eq. (3.10), respectively. To give a fairly accurate graphical representation of the weight function we use one of three numerical methods developed in Ref. [11] for obtaining a good approximation of the density function associated with finite tridiagonal Hamiltonian matrices. As an example, we consider $\rho^\tau(z)$ associated with the orthogonal polynomials $\{Q_n^\tau\}$. Figure (1) shows this density function for a given value of $\mu$ and $\nu$ and for several choices of the parameter $\tau$.

**Appendix C: The Aharonov-Bohm and magnetic monopole**

In this Appendix, we consider briefly the combined Aharonov-Bohm (A-B) effect and a magnetic monopole for which the electromagnetic vector potential $\vec{A}$ has the following components in spherical coordinates

$$A_r = 0,\ A_\theta = 0,\ A_\phi = \frac{a - b\cos\theta}{r\sin\theta}, \tag{C.1}$$

where $a$ and $b$ are real parameters and the A-B magnetic flux strength is $2\pi|a-b|$. The monopole strength is $b$ and it has a singularity along the negative $z$-axis. The wave equation of a charged particle in this electromagnetic potential is obtained by replacing Eq. (2.1) by another with minimal coupling to the vector potential as follows:

$$\left[-\frac{1}{2}\left(\vec{\nabla} - i\zeta\vec{A}\right)^2 + V(\vec{r}) - E\right]\psi = 0, \tag{C.2}$$

where $\zeta = e/\hbar c$. If we also consider the contribution of a static charge $\mathcal{Z}e$ fixed at the origin, then $V(\vec{r}) = \mathcal{Z}/r$ and the total potential field experienced by the charged particle, whose dynamics is described by Eq. (C.2), is non-central and could be written as

$$V(r,\theta) = \frac{\mathcal{Z}}{r} - \frac{\zeta^2 b^2/2}{r^2} + \frac{\zeta}{r^2 \sin^2\theta}\left[\zeta\frac{a^2+b^2}{2} - ma + b(m - \zeta a)\cos\theta\right], \tag{C.3}$$

where we've used the $\phi$-dependence of the complete wavefunction $\psi$ as $e^{im\phi}$ when writing $\partial_\phi \psi = im\psi$. Comparing this potential with (2.3), we conclude that

$$V_r = \frac{\mathcal{Z}}{r} - \frac{\zeta^2 b^2/2}{r^2},\ C_0 = 0,\ \hat{C} = \zeta\left[\zeta(a^2+b^2) - 2ma\right],\ C = 2\zeta b(m - \zeta a), \tag{C.4}$$

Consequently, this situation is compatible with the radial potential configuration given by the case (2.15b), where the centrifugal barrier parameter $\mathcal{B} = -\zeta^2 b^2$ and negative. Using the findings in Sec. 2 of Ref. [1], we can directly obtain the bound states energy spectrum as

$$E_{knm} = -\mathcal{Z}^2/2(k + \nu_{nm} + 1)^2, \tag{C.5}$$

where $v_{nm} = -\frac{1}{2} + \sqrt{\left(\gamma_{nm}^{\pm} + \frac{1}{2}\right)^2 - \zeta^2 b^2}$ and $\gamma_{nm}^{\pm}$ is given by Eq. (3.14). Therefore, bound states for this system exist only if the integers $n$ and $m$ are large enough to satisfy the barrier threshold condition that $\left|\gamma_{nm}^{\pm} + \frac{1}{2}\right| \geq \zeta |b|$. The radial component of the complete bound states wavefunction could also be obtained using results from the same work cited above as

$$R_{knm}(r) = \sqrt{\frac{\lambda_{knm}\Gamma(k+1)}{\Gamma(k+2v_{nm}+1)}} \left(\lambda_{knm}r\right)^{1+v_{nm}} e^{-\lambda_{knm}r/2} L_k^{2v_{nm}}(\lambda_{knm}r), \tag{C.6}$$

where $\lambda_{knm} = 2\sqrt{-2E_{knm}} = 2|\mathcal{Z}|/(k+v_{nm}+1)$. The angular components of the complete wavefunction belong to one of the cases given in Sec. 6. The choice depends on the values of the potential parameters, $a$ and $b$. Pure A-B effect takes place in the absence of the monopole (i.e., for $b = 0$), which also imply zero centrifugal barrier.